\documentclass[
reprint,
superscriptaddress,
 amsmath,amssymb,
 aps,
prl,
]{revtex4-1}

\usepackage{graphicx,color}
\usepackage{dcolumn}
\usepackage{bm}
\begin{document}
\preprint{cond-mat}
\title{Spin and Orbital Contributions to Magnetically Ordered Moments \\ in $5d$ Layered Perovskite Sr$_{2}$IrO$_{4}$}
\author{S.\ Fujiyama}
\email{fujiyama@riken.jp} 
\affiliation{RIKEN, Magnetic Materials Laboratory, Wako 351-0198, Japan}
\author{H.\ Ohsumi}
\affiliation{RIKEN, SPring-8 Center, Sayo, Hyogo 679-5148, Japan}
\author{K.\ Ohashi}
\affiliation{Department of Advanced Materials Science, University of Tokyo, Kashiwa 277-8561, Japan}
\author{D.\ Hirai}
\affiliation{Department of Physics, University of Tokyo, Tokyo 113-0033, Japan}
\author{B.J.\ Kim}
\affiliation{Department of Advanced Materials Science, University of Tokyo, Kashiwa 277-8561, Japan}
\author{T.\ Arima}
\affiliation{RIKEN, SPring-8 Center, Sayo, Hyogo 679-5148, Japan}
\affiliation{Department of Advanced Materials Science, University of Tokyo, Kashiwa 277-8561, Japan}
\author{M.\ Takata}
\affiliation{RIKEN, SPring-8 Center, Sayo, Hyogo 679-5148, Japan}
\author{H.\ Takagi}
\affiliation{Department of Physics, University of Tokyo, Tokyo 113-0033, Japan}
\affiliation{RIKEN, Magnetic Materials Laboratory, Wako 351-0198, Japan}
\date{\today}
\begin{abstract}

The ratio of orbital ($L$) and spin ($S$) contributions to the magnetically ordered moments of a $5d$ transition metal oxide, Sr$_{2}$IrO$_{4}$ was evaluated by non-resonant magnetic x-ray diffraction. We applied a new experimental setting to minimize the error in which we varied only the linear-polarization of incident x-ray at a fixed scattering angle. Strong polarization dependence of the intensity of magnetic diffraction was observed, from which we conclude that the ordered moments contain substantial contribution from the orbital degree of freedom with the ratio of $\langle L\rangle/\langle S\rangle\sim 5.0$, evidencing the pronounce effect of spin-orbit coupling. The obtained ratio is close to but slightly larger than the expected value for the ideal $J_\mathrm{eff}=1/2$ moment of a spin-orbital Mott insulator, $|\langle J_{1/2} | L_{z}| J_{1/2} \rangle|/|\langle J_{1/2} | S_{z}| J_{1/2} \rangle|=4$, which cannot be accounted by the redistribution of orbital components within the $t_{2g}$ bands associated with the elongation of the IrO$_{6}$ octahedra.
\end{abstract} 
\pacs{75.25.-j, 75.70.Tj, 71.30.+}
\maketitle

Magnetic moments of $3d$ transition metal complexes predominantly originate from the spins of the $d$ electrons rather than the orbital momenta because of the quenching of the orbital momenta by the crystal fields. The orbital contribution is treated as a perturbation to the spin, the effect of which is renormalized to the $g$ factor of electrons resulting in a magnetic anisotropy. In heavier elements like $5d$ transition metals, however, the spin-orbit coupling (SOC) is pronounced and drastic change of the electronic states is anticipated. An unconventional Mott state was recently found in a layered perovskite with Ir$^{4+}$ ($5d^{5}$), Sr$_{2}$IrO$_{4}$, where a half-filled band  of $J_\mathrm{eff}=1/2 \; (\mathbf{J}_\mathrm{eff}=\mathbf{S}+\mathbf{L}_\mathrm{eff}=\mathbf{S}-\mathbf{L}$ ) is formed by a strong SOC and a moderate Coulomb repulsion opens a charge gap within the $J_\mathrm{eff}=1/2$ band~\cite{Crawford1994,Kim2008}. Experimental supports for the $J_\mathrm{eff}=1/2$ character of the ordered moments were given by the measurements of XAS and resonant magnetic x-ray scattering (RXS)~\cite{Kim2008,Kim2009}. The strong suppression of the magnetic x-ray scattering at the $L_\mathrm{II}$ edge ($2p_{3/2}\rightarrow 5d$) indicated that the wave function of electrons in charge of magnetism should be reasonably close to that of $J_\mathrm{eff}=1/2$ state. The $J_\mathrm{eff}=1/2$ character of the ordered moments was also identified in other complex iridium oxides~\cite{Kim2012,Fujiyama2012,Fujiyama2012B,Boseggia2012}. 

In reality, even in the prototypical spin-orbital Mott insulator Sr$_{2}$IrO$_{4}$, there should be certain deviation from the ideal $J_\mathrm{eff}=1/2$ state. Factors to destabilize the ionic $J_\mathrm{eff}=1/2$ band could be the followings.

\begin{figure}[htb]
\includegraphics*[width=8cm]{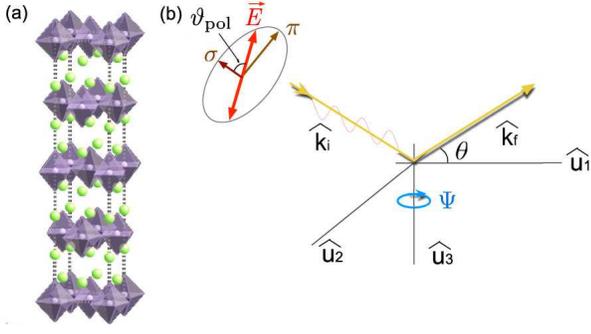}
\caption{(a) Crystal structure of Sr$_{2}$IrO$_{4}$. (b) Schematic diagram of the separation of the orbital and spin contributions to the ordered moments by magnetic x-ray diffraction. The incident x-ray beam is linearly polarized with the angle $\vartheta_\mathrm{pol}$ from $-\hat{u}_{2}$. Here, $k_{i}$ and $k_{f}$ are the incident and scattered wave vectors. $\Psi$ is the azimuth angle that represents the rotation angle of the sample within $\hat{u}_{1}$-$\hat{u}_{2}$ plane.}
\label{fig:scatter}
\end{figure}

\begin{enumerate}
\setlength{\itemsep}{-3pt}
\item The refinement of the crystal structure of Sr$_{2}$IrO$_{4}$ showed an elongation ($\sim 4$ \% along the $c$-axis) and the rotation within a plane of IrO$_{6}$ octahedra~(Fig.~\ref{fig:scatter} (a))~\cite{Crawford1994}. Those should lift the degeneracy of $t_{2g}$.
\item The octahedral ligand field splitting, 10 Dq, is finite, which could give rise to admixture of the $e_{g}$ states~\cite{Watanabe2010,Jin2009,Arita2012}. 
\item The system is close to metallic states and not a strong insulator. Band effect should influence the nature of the ordered moment~\cite{Fujiyama2012B,Cao2007,Carter2012}.
\end{enumerate}

To fully understand the nature of the magnetic moment of a spin-orbital Mott insulator, Sr$_{2}$IrO$_{4}$, estimates of orbital ($\langle L \rangle \equiv \langle \psi|L_{z}| \psi \rangle$) and spin ($\langle S \rangle \equiv \langle \psi|S_{z}| \psi \rangle$) moments of the ordered moment should be the first step to determine the precise wave function, $|\psi \rangle$. The deduced $\langle L \rangle / \langle S \rangle$ ratio would specify the competing source to destabilize the ionic $J_\mathrm{eff}=1/2$ band. Determination of $\langle L \rangle$ and $\langle S \rangle$ was performed by X-ray magnetic circular dichroism (XMCD) technique, which probed only the canted component as the field-induced weak ferromagnetic moments~\cite{Haskel2012}. Our approach to tackle this problem is to probe the orbital and spin magnetizations separately by a polarization analysis of non-resonant magnetic X-ray scattering (NRMXS).

The principle of individual observation of $\langle L \rangle$ and $\langle S \rangle$ by NRMXS was formulated by Blume and Gibbs~\cite{Blume1988}. While spins in the target material can be treated as stational objects like electron charges, the interaction between \textit{orbital motion} of electrons around nuclei and incident x-ray is expected to include a relativistic process.  The scattering amplitude at a wave vector $\mathbf{K}$ of magnetic scattering is a summation of spin and orbital contributions as
\begin{equation}
\langle M_{m} \rangle =-i \tau [\mathbf{S(K)}\cdot \mathbf{B}+\mathbf{L(K)}\cdot \mathbf{B_{0}}],
\end{equation}
where 
\begin{equation}
\mathbf{B}=(\epsilon' \times \epsilon)-(\hat{k_\mathrm{f}}\times \epsilon')\times(\hat{k_\mathrm{i}}\times \epsilon)+(\hat{k_\mathrm{f}}\cdot \epsilon)(\hat{k_\mathrm{f}}\times \epsilon')-(\hat{k_\mathrm{i}}\cdot \epsilon')(\hat{k_\mathrm{i}}\times \epsilon)
\end{equation}
and
\begin{equation}
\mathbf{B_{0}}=(\epsilon ' \times \epsilon).
\end{equation}
The $\hat{k_\mathrm{i}} (\hat{k_\mathrm{f}})$ and $\epsilon (\epsilon')$ denote the unit vectors pointing the propagation and the polarization of the incident (scattered) x-ray, as shown in Fig.~\ref{fig:scatter} (b). It should be stressed that spin and orbital densities ($\mathbf{S(K)}$ and $\mathbf{L(K)}$) are coupled with different matrices. This is a crucial advantage of magnetic x-ray diffraction over neutron diffraction in identifying the orbital and spin contributions. The magnetic scattering matrix of neutrons is simply written as
$\langle M_{n}\rangle=\left[ \frac{1}{2}\mathbf{L(K)+S(K)} \right] \cdot \mathbf{C}$. Here, $\left[\frac{1}{2}\mathbf{L}+\mathbf{S}\right]$ couples to a matrix $\mathbf{C}=\hat{K}\times(\boldsymbol \sigma \times \hat{K})$, using $\hat{K}$, the unit vector along $\mathbf{K}$, and neutron spin operator $\boldsymbol \sigma$. For this reason, one cannot separate the orbital and spin contributions to the ordered moments by neutron scattering.

In this Letter, we report the ratio of orbital and spin contributions to the magnetically ordered moments of Sr$_{2}$IrO$_{4}$, by utilizing the merit of the magnetic scattering of x-ray. We applied a new experimental setting of NRMXS with fixed $\mathbf{K}$ and azimuthal angle, $\Psi$, which minimizes the experimental error in estimating the $\langle L\rangle /  \langle S\rangle$ ratio~\cite{Ohsumi2004}. The ratio
$\langle L \rangle/ \langle S \rangle\equiv |\langle \psi | L_{Z}| \psi \rangle | / | \langle \psi  | S_{Z}| \psi  \rangle |$ is estimated to be $5.0 \pm 0.7$ at a magnetic Bragg reflection $\mathbf{K}=$(1 0 22). The obtained ratio is close to the expected value for the $J_\mathrm{eff}=1/2 \, (J_{1/2})$ state at $\mathbf{K}=0$, $|\langle J_{1/2}|L_{z}|J_{1/2} \rangle|/|\langle J_{1/2}|S_{z}|J_{1/2} \rangle|=4$, but certainly shows an enhancement of the orbital component. The deviation from the ideal $J_\mathrm{eff}=1/2$ cannot be explained by the admixture of the $J_\mathrm{eff}=3/2$ state associated with the elongation of an IrO$_{6}$ octahedron. We also demonstrate the orientation of the magnetic moments by rotating the azimuthal angle ($\Psi$) with 1$^{\circ}$ accuracy.

A single crystal of Sr$_{2}$IrO$_{4}$ with a dimension of 1 mm$\times $0.3 mm $\times$0.1 mm was synthesized by a flux method. Antiferromagnetic long-range order below $T_\mathrm{N}=230$ K was confirmed by magnetization and RXS~\cite{Kim2009,Fujiyama2012}. NRMXS measurements were performed at 10 K, where the magnitudes of the ordered moments were almost saturated upon cooling. We used BL19LXU at SPring-8 with a purely polarized x-ray~\cite{Yabashi2001}, which enabled us to control the linear polarization of the incident x-ray precisely.

The configuration of NRMXS measurement is shown in Fig.~\ref{fig:scatter} (b). Hereafter we use a Cartesian coordinate system, defined by unit vectors $\hat{u}_{1}$, $\hat{u}_{2}$, and $\hat{u}_{3}$, which point to the direction of $\hat{k}_{i}+\hat{k}_{f}$, $\hat{k}_{i} \times \hat{k}_{f}$, and $\hat{k}_{i}-\hat{k}_{f}$, respectively. The origin of the azimuth angle, $\Psi$ is set to the crystalline $b$-axis $\parallel \hat{u}_{1}$ direction. Each experimental data set was taken by rotating the linear polarization direction $\vartheta_\mathrm{pol}$ with fixed $\theta$ and $\Psi$, and fit by the theoretical curve with a  parameter $\langle L\rangle/\langle S \rangle$. The ratios $\langle L \rangle/ \langle S \rangle$ were determined for three sets of azimuth angles, $\Psi$, to confirm the accuracy of our estimate. The matrix form of NRMXS is given in Ref.~\onlinecite{Blume1988} as
\begin{widetext}
\begin{eqnarray}
\langle M_{m}\rangle= \left(
\begin{array}{cc}
 S_{2} \sin 2 \theta & -2\sin^{2}\theta [ ( L_{1}+S_{1})\cos \theta  - S_{3} \sin \theta  ] \\ 2\sin^{2} \theta [ (L_{1}+S_{1}) \cos \theta+ S_{3} \sin \theta] & \sin 2\theta [2L_{2} \sin^{2}\theta +S_{2}]\\
\end{array} \right).
\label{eq:mu}
\end{eqnarray}
\end{widetext}
Here, $S_{j}(L_{j})$ is the direction cosine of spin (orbital) contribution to the ordered moment along $\hat{u}_{j}$. When the incident x-ray is linearly polarized with an angle $\vartheta_\mathrm{pol}$, the expected intensity of magnetic diffraction is expressed as $I_{m} \propto \mu_{\sigma'}^{2}+\mu_{\pi'}^{2}$, where
\begin{eqnarray}
\left(
\begin{array}{cc}
\mu_{\sigma'} \\ \mu_{\pi'}
\end{array}
\right)
=\langle M_{m} \rangle
\left(
\begin{array}{cc}
\cos \vartheta_\mathrm{pol} \\ \sin \vartheta_\mathrm{pol}
\end{array}
\right).
\label{eq:matrix}
\end{eqnarray}
$I_{m}$ is expected to follow a sinusoidal $\vartheta_\mathrm{pol}$ dependence for non-zero orbital contribution to the magnetic moments. 

Previously, Eq.~(\ref{eq:mu}) was applied to magnetic x-ray scattering for holmium or uranium compounds to evaluate $\langle L \rangle/\langle S \rangle$, where the $\vartheta_\mathrm{pol}$ value was set at $0$ or $\pi/2$~\cite{Gibbs1988,Gibbs1991,Langridge1997}. The scattering-angle ($\theta$) dependent ratio of $\pi'$ and $\sigma'$ detections,
\begin{equation}
I_{\pi'}/I_{\sigma'}=\sin^{2} \left( \frac{\theta}{2} \right)\frac{\left[ 2L_{1}(\mathbf{K})+S_{1}(\mathbf{K})-S_{3}(\mathbf{K})\tan \left( \theta/2 \right) \right]^{2}}{S_{2}(\mathbf{K})^{2}}
\end{equation}
yielded a ratio of orbital and spin contributions of the magnetic moments. However, this approach using an analyzer crystal for the polarization analysis was subject to a large experimental error due to the insufficient intensities of diffraction. Besides, the magnetic moments of holmium and of uranium compounds are well-localized $f$ electrons where $\langle L (\mathbf{K})\rangle$ and $\langle S (\mathbf{K}) \rangle$ have small $\theta$ dependences. In case of 5d electrons, the $\theta$ dependence could be much more significant and the measurement of the $\mathbf{K}$ dependent $I_{\pi '}/I_{\sigma '}$ would cause a large error. An experimental approach without an analyzer crystal that is precise enough to estimate $\langle L \rangle/\langle S \rangle$ ratio even at a fixed-$\mathbf{K}$ has been anticipated, which is particularly crucial for $5d$ transition metal oxides. We examined the $\vartheta_\mathrm{pol}$ dependence of the diffraction with fixed $\mathbf{K}$, by controlling two $\pi/2$ phase plates of diamond to realize $\Delta\vartheta_\mathrm{pol}\sim1^{\circ}$~\cite{Ohsumi2004}.

We measured the $\vartheta_\mathrm{pol}$-dependent intensity of (1 0 22) magnetic reflection. The photon energy of incident x-ray was set to 10.45 keV. The  scattering angle was $\theta=31.1^{\circ}$ corresponding to the (1 0 22) magnetic Bragg reflection. The experimental configuration with fixed diffraction angle ($\theta$) and azimuth angle ($\Psi$) maintained the irradiated area on the surface unchanged.
\begin{figure}[htb]
\includegraphics*[width=5cm]{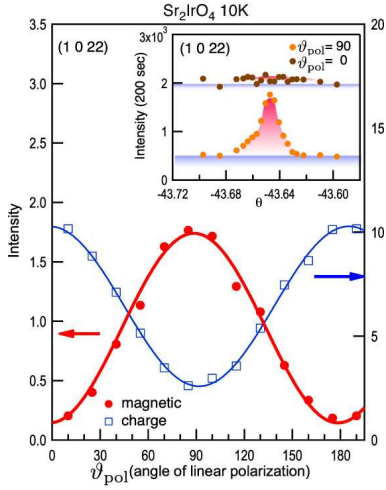}
\caption{$\vartheta_\mathrm{pol}$ dependence of magnetic and charge scatterings in the antiferromagnetic state for $\Psi=90^{\circ}$. The diffractions for $\vartheta_\mathrm{pol}=0^{\circ}$ and $90^{\circ}$ are shown in the inset.}
\label{fig:LS1}
\end{figure}

\begin{figure}[htb]
\includegraphics*[width=8cm]{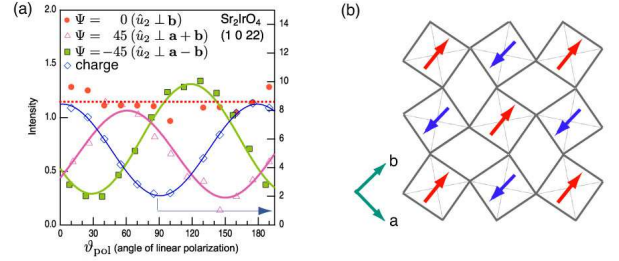}
\caption{(a) The $\vartheta_\mathrm{pol}$ dependence of magnetic and charge scattering for three azimuth angles. The charge scattering did not depend on $\Psi$. The solid and dashed lines represent fitted curves for three azimuthal angles with a fixed $\langle L \rangle/ \langle S \rangle=5.0$. It should be noted that no fitting parameter other than the scaling factor is used. The phases of the sinusoidal $\vartheta_\mathrm{pol}$ dependence are well reproduced by Eq.~(\ref{eq:mu}).
 (b) Magnetic structure of the ordered state in Sr$_{2}$IrO$_{4}$ determined by the measurements of the azimuth angle ($\Psi$) dependent magnetic diffraction.}
\label{fig:Ha}
\end{figure}

The obtained $\vartheta_\mathrm{pol}$ dependence of charge and magnetic scatterings for $\Psi=90^{\circ}$ ($a \perp \hat{u}_{2}$) is shown in Fig.~\ref{fig:LS1}. Charge scattering is expressed as
$I_{c} \propto \gamma_{c,\sigma'}^{2}+\gamma_{c,\pi'}^{2}$ with
\begin{equation}
\left(
\begin{array}{cc}
\gamma_{c,\sigma'} \\  \gamma_{c,\pi'}
\end{array}
\right)
= \rho(\mathbf{K}) \left( 
\begin{array}{cc}
1 & 0 \\
0 & \cos 2\theta \\
\end{array}
\right)\left(
\begin{array}{c}
\cos \vartheta_\mathrm{pol} \\
 \sin \vartheta_\mathrm{pol}
\end{array} \right).
\end{equation}
Both the charge and magnetic diffraction intensities show sinusoidal $\vartheta_\mathrm{pol}$ dependence. Large amplitude of the sinusoidal curve for the magnetic scattering evidences considerable contribution of the orbital component to the ordered moments. The antiphase $\vartheta_\mathrm{pol}$ dependence to that of the charge scattering indicates $b$-collinear antiferromagnetic moments ($\hat{u}_{2} \parallel \mathbf{M}$). We rotated the sample around the $Q$ vector, and measured the magnetic diffraction for other azimuthal angles, $\Psi=0^{\circ}, \pm 45^{\circ}$ corresponding to $\hat{u}_{2} \perp \bm{b}, \hat{u}_{2}\perp \bm{a}  \pm \bm{b}$, respectively, as shown in Fig.~\ref{fig:Ha} (a). One prominent finding is that $I_{m}$ becomes nearly independent of $\vartheta_\mathrm{pol}$ for $\Psi =0$, which we could obtain only in the case that the magnetic moments lie along the $\hat{u}_{1}$ direction. This firmly establishes that the magnetic moments lie along the $b$-axis, and combined with the antiparallel alignment of moments by the RXS measurements~\cite{Kim2009}, we determined the magnetic structure as shown in Fig.~\ref{fig:Ha} (b). By fitting the data with Eq.~(\ref{eq:mu}) by taking $\mathbf{M} \parallel b$ into account, we estimate the ratio of orbital and spin contributions to the ordered moment $\langle L \rangle/ \langle S \rangle$ as $5.0 \pm 0.7$. The ratios obtained by other azimuth angles are consistent with the value $\langle L \rangle/ \langle S \rangle \sim 5.0$ as shown in Fig.~\ref{fig:Ha} (a). These results confirm the vital role of the large spin-orbit coupling.

Considering the ionic wave function with the magnetic moment along the $b$-axis, the $J_\mathrm{eff}=1/2$ state is written by choosing the quantized axis $\bm{Z} \equiv  1/\sqrt{2} \left( \bm{x}+\bm{y} \right)$ as\\
\begin{eqnarray*}
|J_\mathrm{eff}  & = & 1/2,\pm \rangle =\left( 1/\sqrt{6} \right) \left[  \left( |yz,+\rangle \pm |yz,-\rangle \right) \right. \\ && \left. \pm \left( |zx,+ \rangle \mp |zx,- \rangle \right) \mp \sqrt{2}i |xy, \mp \rangle \right] \\ &=& \left( i/2 \sqrt{3} \right) |R_{52} \rangle\left[  \left(1\pm i \right)\left(|Y_{2}^{1},+\rangle \pm |Y_{2}^{-1},- \rangle \right) \right. \\&& \left. \pm \left(1\mp i \right) \left(|Y_{2}^{1},-\rangle \pm |Y_{2}^{-1},+\rangle \right) \pm \sqrt{2}i |Y_{2}^{2}-Y_{2}^{-2}, \mp \rangle\right].
\end{eqnarray*}
Here, $| R_{52} \rangle$ and $| Y_{2}^{l_{z}} \rangle$ are the radial and spherical wave functions of $5d$ electrons respectively, and the $\pm$ signs denote up and down spins quantized to the $\bm{b}$ direction. Using this $J_\mathrm{eff}=1/2$ wave function,
 the ratio of the expectation values of $L_{Z}=1/\sqrt{2} \left( L_{x}+L_{y} \right)$ and $S_{Z}=1 / \sqrt{2} \left( S_{x}+S_{y} \right) $ is $|\langle J_{1/2} |L_{Z}| J_{1/2} \rangle |/| \langle J_{1/2} |S_{Z}| J_{1/2} \rangle|=4$ at $\mathbf{K}=0$.  The present measurements were performed at $\mathbf{K}=$(1 0 22) ($\theta=31.1^{\circ}$) and the expected ratio is shifted by spin and orbital form factors of $d$-electrons, which are composed of Fourier transforms of three $t_{2g}$ wave functions. The calculated form factors defined by $f_{1(2)}(\mathbf{K})\equiv\oint |R_{52}|^{2}|Y_{2}^{1(2)}|^{2}\exp(-2\pi i (\mathbf{K}\cdot \mathbf{r})) d\mathbf{r}$ are 0.636 for $l_{z}=1(|yz\rangle,|zx\rangle)$ and 0.873 for $l_{z}=2(|xy\rangle)$ spherical wave functions. Using the form factors at  $\mathbf{K}$ = (1 0 22), the expected ratio of orbital and spin contributions is shifted to $\left[1/3\cdot  (f_{1}+f_{2}) \right]/ \left[1/6 \cdot f_{2} \right]=3.5$~\cite{RatioCalc}, that is slightly smaller than 4, the value at $\mathbf{K}=0$. The observed ratio of orbital and spin contributions to the moment, $\langle L \rangle/ \langle S \rangle \sim 5.0$ is not far away from 3.5, showing that the wave function of Sr$_{2}$IrO$_{4}$ is approximated as the $J_\mathrm{eff}=1/2$ function. This is in support of the dominant effect of the spin-orbit coupling.
 
Since the expected ratio for the ideal $J_\mathrm{eff}=1/2$ is $|\langle J_{1/2} |L_{Z}| J_{1/2} \rangle |/| \langle J_{1/2} |S_{Z}| J_{1/2} \rangle|<4$ because of $f_{1}<f_{2}$ for any $\mathbf{K}$, the observed $\sim 40 $ \% overshoot of $\langle L \rangle/ \langle S \rangle$ is not an experimental error but shows a physically significant hybridization of the $J_\mathrm{eff}=1/2$ with other states. We discuss possible sources for this enhancement of the orbital contribution.

The crystal structure analysis shows a $\sim 4$ \% elongation of IrO$_{6}$ octahedra along the $c$-axis, which causes a ligand field splitting of the triply degenerated $t_{2g}$ manifold into $|xy \rangle$ with upper energy level and doubly degenerated $|yz\rangle$ and $|zx\rangle$ orbitals. Considering only $|xy\rangle$, $|yz \rangle$, and $|zx \rangle$, the wave function in the tetragonal field can be described as \\
\begin{eqnarray*}
|\psi \rangle & = & 1/\sqrt{6} \left[  \left( 1- \delta \right) \left( |yz,+\rangle \pm |yz,-\rangle \right) \right. \\ & & \left.  \pm \left( 1- \delta \right) \left( |zx,+ \rangle  \mp |zx,- \rangle \right) \mp \sqrt{2}i \left( 1+2\delta \right) |xy, \mp \rangle \right].
\end{eqnarray*}
To justify the observed $\langle L \rangle / \langle S \rangle$ ratio at $\mathbf{K}$, $|\langle \psi |L_{Z}|\psi\rangle|/|\langle \psi |S_{Z}|\psi\rangle|=5.0$, we obtain $\delta=-0.11$, which implies the reduction of the $|xy \rangle$ orbital with a down spin. The elongated IrO$_{6}$ octahedra along the $c$-axis should enhance $|xy\rangle$ character resulting in a positive value of $\delta$, which is opposite to the observed coefficient. The intra-$t_{2g}$ redistribution of orbital states caused by the elongation of an IrO$_{6}$ octahedron is clearly insufficient to understand the enhancement of the orbital contribution. Other hybridization effect should be invoked. For example, the staggered in-plane rotation of IrO$_{6}$ octahedra could induce the hybridization of $e_{g}$ and $t_{2g}$ states due to the finite 10 Dq by the cubic crystal field. The weak Mott character of electrons proximity to a bad metal can also destabilize the ionic picture.

To conclude, we present a quantitative estimate of the ratio of orbital and spin contributions to the magnetically ordered moment of $5d$ based antiferromagnet, Sr$_{2}$IrO$_{4}$, utilizing NRMXS. The experimental setting applied to this material offers a unique opportunity to unveil the real electronic state produced by the novel interplay of the strong spin-orbit coupling, lattice distortion, and hybridizations with other orbital states. The magnetic moment is dominated by orbital angular momentum with the value $\langle L \rangle/ \langle S \rangle\sim 5.0 \pm 0.7$, which is not far from the $| J_\mathrm{eff}=1/2\rangle$ spin-orbital Mott picture for this material. A nearly 40 \% enhancement of the orbital contribution from the ideal $J_\mathrm{eff}=1/2$ state was observed. This cannot be ascribed to an intra-$t_{2g}$ redistribution of the orbital component associated with the elongated IrO$_{6}$ octahedron but clearly points the importance of the hybridization with other orbital states. We have also succeeded in determining the magnetic structure including both the wave vector and the direction of ordered moments by analyzing the polarization dependence of NRMXS.

\begin{acknowledgments}
We are grateful to T.\ Shirakawa, H.\ Watanabe, S.\ Yunoki, T.\ Komesu, S.\ Sakai and N.\ Shannon for their technical supports and fruitful discussions. The synchrotron radiation experiments were performed at BL19LXU in SPring-8 with the approval of RIKEN (20080047). This work was supported by Grant-in-Aid for Scientific Research (S) (24224010) from JSPS.
\end{acknowledgments}



\begin{thebibliography}{}
\bibitem{Crawford1994}M.K.\ Crawfard \textit{et al.} Phys.\ Rev.\ B \textbf{49}, 9198 (1994).
\bibitem{Kim2008}B.J.\ Kim, Hosub Jin, S.J.\ Moon, J.-Y.\ Kim, B.-G.\ Park, C.S.\ Leem.\ Jaejun Yu, T.W.\ Noh, C.\ Kim, S.-J.\ Oh, J.-H.\ Park, V.\ Durairaj, G.\ Cao, and E.\ Rotenberg, Phys.\ Rev.\ Lett.\ \textbf{101}, 076402 (2008).
\bibitem{Kim2009}B.J.\ Kim, H.\ Ohsumi, T.\ Komesu, S.\ Sakai, T.\ Morita, H.\ Takagi, and T.\ Arima, Science \textbf{323} 1329 (2009).
\bibitem{Kim2012} J.W.\ Kim, Y.\ Choi, Jungho Kim, J.F.\ Mitchell, G.\ Jackeli, M.\ Daghofer, J.\ van den Brink, G.\ Khaliullin, and B.J.\ Kim, Phys.\ Rev.\ Lett. \textbf{109}, 037204 (2012).
\bibitem{Fujiyama2012}S.\ Fujiyama, H.\ Ohsumi, T.\ Komesu, J.\ Matsuno, B.\ J.\ Kim, M.\ Takata, T.\ Arima, and H.\ Takagi, Phys.\ Rev.\ Lett.\ \textbf{108}, 247212 (2012).
\bibitem{Fujiyama2012B}S.\ Fujiyama, K.\ Ohashi, H.\ Ohsumi, K.\ Sugimoto, T.\ Takayama, T.\ Komesu, M.\ Takata, T.\ Arima, and H.\ Takagi, Phys. Rev. B, \textbf{86}, 174414, (2012).
\bibitem{Boseggia2012} S.\ Boseggia, R.\ Springell, H.C.\ Walker, A.T.\ Boothroyd, D.\ Prabhakaran, D.\ Wermeille, L.\ Bouchenoire, S.P.\ Collins, and D.F.\ McMorrow, Phys.\ Rev.\ B \textbf{85}, 184432 (2012).
\bibitem{Watanabe2010}H.\ Watanabe, T.\ Shirakawa, and S.\ Yunoki, Phys.\ Rev.\ Lett.\ \textbf{105}, 216410 (2010), and private communications.
\bibitem{Jin2009}H.\ Jin, H.\ Jeong, T.\ Ozaki, and J.\ Yu, Phys.\ Rev.\ B \textbf{80}, 075112 (2009).
\bibitem{Arita2012}R.\ Arita, J.\ Kune\v{s}, A.V.\ Kozhevnikov, A.G.\ Eguiluz, and M.\ Imada, Phys.\ Rev.\ Lett. \textbf{108}, 086403 (2012).
\bibitem{Cao2007}G.\ Cao, V.\ Durairaj, S.\ Chikara, L.E.\ DeLong, S.\ Parkin, and P.\ Schlottmann, Phys.\ Rev.\ B, \textbf{76}, 100402 (2007).
\bibitem{Carter2012}Jean-Michel Carter, V.V.\ Shankar, M.A.\ Zeb, and Hae-Young Kee, Phys. Rev. B, \textbf{85}, 115105 (2012).
\bibitem{Haskel2012} D.\ Haskel, G.\ Fabbris, Mikhail Zhernenkov, P.\ P.\ Kong, C.\ Q.\ Jin, G.\ Cao, and M.\ van Veenendaal, Phys.\ Rev.\ Lett.\ \textbf{109}, 027204 (2012).
\bibitem{Blume1988}M.\ Blume and D.\ Gibbs, Phys.\ Rev.\ B, \textbf{37} 1779 (1988).
\bibitem{Ohsumi2004}H.\ Ohsumi, M.\ Mizumaki, S.\ Kimura, M.\ Takata, and H.\ Suematsu, Physica B, \textbf{345}, 258 (2004).
\bibitem{Yabashi2001}M. Yabashi, \textit{et al.} Nucl. Instr. Meth. Phys. Rev. A \textbf{467-468}, 678 (2001).
\bibitem{Gibbs1988}D.\ Gibbs, E.R.\ Harshman, E.D.\ Isaacs, D.B.\ McWhan, D.\ Mills, and V.\ Vettier, Phys.\ Rev.\ Lett.\ \textbf{61}, 1241 (1988).
\bibitem{Gibbs1991}D.\ Gibbs, G.\ Grubel, D.R.\ Harshman, E.D.\ Issacs, D.B.\ McWhan, D.\ Mills, and C.\ Vettier, Phys.\ Rev.\ B \textbf{43}, 5663 (1991).
\bibitem{Langridge1997}S.\ Langridge and G.\ H.\ Lander, N.\ Bernhoeft, A.\ Stunault, C.\ Vettier, G.\ Gr\"{u}bel, and C.\ Sutter, F.\ de Bergevin, W.\ J.\ Nuttall and W.\ G.\ Stirling, K.\ Mattenberger and O.\ Vogt, Phys.\ Rev.\ B, \textbf{55} 6392 (1997).
\bibitem{RatioCalc}The expectation values of $L_{Z}$ and $S_{Z}$ are expressed using the spherical wave functions as
\begin{eqnarray*} \langle J_{1/2,+} | L_{Z} |J_{1/2,+} \rangle=1/6 \langle R_{52}|R_{52} \rangle \left[ \langle Y^{2},-|Y^{2},-\rangle \right. \\  \left. + \langle Y^{-2},-|Y^{-2},- \rangle+\langle Y^{1},-|Y^{1},- \rangle+\langle Y^{-1},-|Y^{-1},- \rangle\right] \end{eqnarray*}
and $\langle J_{1/2,+} | S_{Z} |J_{1/2,+} \rangle=1/6 \langle R_{52}|R_{52} \rangle \langle Y^{2},- | Y^{2},-\rangle$.
\end{thebibliography}
\end{document}